\documentstyle[aps]{revtex}
\begin{document}
\draft
\title{Wave Function Interpretation and Quantum Mechanics Equations}
\author{Andrey V. Novikov-Borodin $^{*}$}
\address{ Institute for Nuclear Research, Russian Academy of Sciences \\
117312 Moscow, 60-th October Anniversary prospect, 7a. Russia}
\date{\today}
\maketitle
\begin{abstract}
The quantum mechanics description of a physical object stretched in space and 
stable in time from the relativistic space-time properties point of view, 
introduced in special theory of relativity, is considered and analysed. 
The mathematical model of physical objects is proposed. This model 
gives a possibility to unite a description of corpuscular and wave properties
of real physical objects, i.e. fields and particles. There are substantiated 
an approach and a mathematical pattern which give a possibility to describe 
physical object not only in causal, but also in absolute remote fields 
of the Minkowski space. 
Applying the proposed approach to the microcosm description, one can get 
the equations that in passage to the limit transfer to such quantum mechanics 
equations as Schr\H{o}dinger, Klein-Gordon-Fock and in particular 
case - the wave equation. 
The event nature of the received equations is discussed. It is shown that all 
mentioned equations reflect the space-time relativistic properties during the 
description of the invariant and non-invariant physics object characteristics. 
\end{abstract}
\thanks{$^{*}$ E-mail address: Novikov@al20.inr.troitsk.ru}

% body of paper here
\section*{Introduction}

The predictions brilliantly proved in experiments won for the quantum theory 
the reputation of one of the most successful physical theories. However, up 
to present, the disputes about its meaning and limits of its implementation are 
not quiet yet. This is an unique phenomenon in the history of science [1,2].
 The Nobel Prize laureate in physics M. Gell-Mann characterised the quantum 
physics as a discipline ``full of mysteries and paradoxes, that we do not 
completely understand but are able to use. As we know, it perfectly operates 
in the physics reality description, but as sociologists would say, it is 
anti-intuitive discipline. The quantum physics is not a theory, but limits, 
in which, as we suppose, any correct theory needs to be included'' [3]. \\
The logistic analysis of a quantum mechanics as a science leads to the 
conclusion about its incompleteness and non-completability, in consequence 
of an inconsistency of the quantum objects, that is fixed in a corpuscular-
wave dualism [4]. The theory incompleteness is an original ``payment'' for 
a tendency to create a non-contradictory description of contradictory objects. 
One of consequences of the quantum mechanics logistic analysis is a proof of 
an absence of a positive decision for a hidden parameter method, from point 
of view that ``there are no any possibilities for more complete description 
in standard quantum mechanics theory limits. Its realization needs a quantum 
mechanics creation on a principal different basis''. \\
The facts of the particles and anti-particles annihilation with the photon 
and neutrino creation, the birth of particles from different classes during 
the high energy photons interactions are the circumstantial evidence of the 
unified origin of fields and particles. \\
Do not discussing about ``a forced formalism'', i.e. a science penetration 
into the fields with different from ``everyday'', principal new forms and 
meanings, it would be desirable to decrease the formal apparatus as possible, 
by changing it to the system of views. \\
From this point of view, the development of physical object models, which 
give a possibility to unite the description of corpuscular and wave properties 
of real objects, i.e. fields and particles, has a conceptual meaning. Models 
need to correlate with the existing quantum mechanics approach, so as it has a 
brilliant experimental confirmation. Moreover, although some quantum mechanics 
postulates and concepts should be a corollary from the properties of the 
proposed model and properties of the space, where the object exists. If the 
relativistic properties are included in the space description, the 
corresponding expressions describing the object behaviour should have 
a relativistic nature and in passage to the limit should be transfer to 
well-known quantum mechanics equations. At any probability, if Minkowski 
space is examined, the absolute remote fields have to be included in the 
physical object description. \\
An effort has been undertaken to solve mentioned problems and ``to pave 
a way'' for the complement the existing quantum mechanics theory for the full, 
correct, non-contradictory theory, dreamed by Louis de Broglie, mentioned by 
Marry Gell-Mann, looked for by Ilya Prigogine. \\

\section{Physical object}
\label{sec:level1}

In special theory of relativity the object evolution is examined in a 
four-dimension space-time continuity with a pseudo-Euclidean metric. The 
supposition about the space-time continuity strikes against the conceptual 
problem, that we will call the scale problem. Let's consider the four-
dimensional space-time as some mathematical set. The Cantor's power of 
a neighbourhood of any space point is equal to the power of whole space. 
The question is: what defines the observed size of a physics object, for 
example, elementary particles? The size of the observer could not be a reason. 
Moving this way we will come to the dilemma about the initial appearance of 
a hen or an egg. The speed of light as a fundamental constant connects
space and time measurements but, apparently, is not good for a role of the 
scale coefficient. In all probability the problem can be solved by introducing 
in a description the discreteness. Considering a time as some parameter 
characterising the changing of processes in space and putting these processes 
in order during the analyses in a chronological sequence, realising the 
causality principle, the hypothesis about the space discreteness can be 
introduced. This way it is not difficult to demand the time continuity as 
some abstract parameter. However, of course, it is not necessary to refuse 
completely from the possibility of the time discreteness. Whether or not the 
discreteness is necessary. Let's consider the possible variants of putting 
the events in order in inertial frames of reference. We will leave
the question of the space discreteness open. \\
A conventional approach is the following [5]. Some frame is introduced in the 
four-dimensional continuity, which is represent a set of four continuous 
marks $(x,y,z,\tau)=(\vec{r},\tau)$  over the space and time coordinates. 
It is established, that the infinite set of the equivalent frames exists, 
and these frames are connected to each other with the help of four continued 
differentiable 
functions with the non-zero functional determinant. Usually, this demand is 
connected to the principle of the equivalence of inertial frames of 
references. It is considered that some property, unchangeable with these 
transformations, can correspond to each point. The property, expressed by the 
number, that ``by order'' don't change during the transformations of the 
frames of references, is called invariant or scalar. It is said about an 
invariant or scalar field if this correspondence takes place not only for 
one concrete point but some number is compared with every point from some 
defined region, and all these numbers reflect the same invariant property. 
Thus, the scalar field is defined by the function against coordinates 
$\phi(\vec{r},\tau)$, that can be interpreted as some continuous physical 
object. For the stable in time physical object, due to the invariance of values 
in its every point, the operation integral can be composed for every point
and by the principle of the minimum effect the trajectory of the object 
motion may be defined. This trajectory will represent the chain of the 
events putting in order in time. We will consider this approach as the 
classical one. \\
It often needs to have deal with non-invariant characteristics of 
physical objects. Let's pose a set of one or more numbers 
$(g(\vec{r},\tau),h(\vec{r},\tau), \dots)$ in some inertial frame of 
references, each of that is not a scalar. If there is a functional on 
one or more elements of this set, that is a scalar or a scalar field, 
the process of putting events in order may be done, although on an 
indirect way or with some limitations. We will call such sets of numbers 
functional, also as the corresponding fields. \\
Let's introduce a point functional, that is important for us. Further 
we will call it the normalised functional. Let's combine the function 
$f(\vec{r},\tau)=g(\vec{r},\tau)+i h(\vec{r},\tau)$, where $i$ is an 
imaginary unit. The corresponding scalar field we will define as follows:
\begin{equation}
\phi(\vec{r},\tau)=|f(\vec{r},\tau)|^2=g(\vec{r},\tau)^2+h(\vec{r},\tau)^2 .
\label{n1}
\end{equation}
Note, that the function $f(\vec{r},\tau)$ defines the normalised functional 
field. \\
The invariant coordinate transformations in different frames of references 
in special theory of relativity are given by Lorentz transformations: 
\begin{equation}
x'=x,~y'=y,~z'=\gamma(z-\beta\tau),~\tau'=\gamma(\tau-\beta z),
\label{n2}
\end{equation}
where $\beta=V/c$, $\gamma=1/\sqrt{1-\beta^2}$, $V$ is a velocity of the 
frame of references $\bf K$ related to a fundamental frame of references 
$\bf K'$, and, without loosing the generality, we will consider that the 
velocity vector is parallel to axes $0z$ and $0z'$. Parameters in the 
fundamental frame will be marked by apostrophes. \\
Let the function $g'(\vec{r'},\tau')$ in some field $\Omega$ of the Minkowski 
space defines the functional field, corresponding to the physical object 
in its fundamental frame $\bf K'$, extensive in space and stable in time. 
Let's mark as $V'$ the cross-section of $\Omega$ in $\bf K'$ by the 
hyper-plane $\tau'= const$. Note, that at any fixed moment of time the 
interval between any two events from $V'$ will be space-like, because 
this hyper-plane will lay down in absolute remote fields of Minkowski space. 
For any point  from $V'$, by virtue of physical object stability, it will 
be such time interval $\Delta\tau'$, what 
$g'(\vec{r'},\tau')=g'(\vec{r'},\tau'+\Delta\tau')$. 
Thus, the function $g'(\vec{r'},\tau')$ will describe the stationary process, 
defined in the space point $\vec{r'}$ (or in some discrete field of space). 
If the time average $<|g'|>_T$, $<|g'|^2>_T$  exist with $T\rightarrow{\infty}$,
$\left( <g(t)>_T ={\int^{T}_{-T} g(t) dt}/(2T) \right)$ and $g'(\vec{r'},\tau')$ 
is the limited variation function on each finite interval of $\tau'$, it can
be represented as a sum of the average value $<g'>_T =g'_0 (\vec{r'})$,
some number of periodical components and non-periodical component $g'_a$ [6]:
\begin{equation}
g'(\vec{r'},\tau')=g'_0 (\vec{r'})+\sum^{\infty}_{k=1} g'_k (\vec{r'})
\cos (\omega'_k \tau'+\alpha'_k )+g'_a (\vec{r'},\tau').
\label{n3}
\end{equation}
Considering $g'(\vec{r'},\tau')$ as a part of the functional field (1),
we will define this field as:
\begin{equation}
\psi'(\vec{r'},\tau')=\sum^{\infty}_{k=0} q'_k (\vec{r'})
\exp (i\omega'_k \tau')+q'_a (\vec{r'},\tau').
\label{n4}
\end{equation}
where $\omega'_0=0$, the limit of $<\psi'\exp(i\omega\tau')>_T$ with
$T\rightarrow \infty$ is equal to $q'_k (\vec{r'})=g'_k (\vec{r'}) 
\exp(i\alpha'_k (\vec{r'}))$ with $\omega=\omega'_k$, $(k = 0,1,2, \dots)$,
and is equal to zero for all other values of $\omega$. The analogous
limit for the non-periodical component always is equal to zero for any
$\vec{r'}$ from $V'$. Functions $g'_k (\vec{r'})$ and $\alpha'_k (\vec{r'})$
are real.\\
In frame $\bf K$ the considered function will be defined as
$\psi(\vec{r},\tau)=\psi'(\vec{r'}(\vec{r},\tau),\tau'(\vec{r},\tau))$,
that, with taking into account (2) and, designating $\xi=\gamma(z-\beta\tau)$,
$\eta=\gamma(\tau-\beta z)-\tau$, may be represented as follows:
\begin{equation}
\psi(\vec{r},\tau)=\sum^{\infty}_{k=0}q_k(x,y,\xi)
\exp[i\omega'_k (\eta+\tau)]+q_a(\vec{r},\tau).
\label{n5}
\end{equation}
For every fixed harmonics (5) the scalar field (1) corresponding to the
introduced normalised one will be equal to squared amplitude of the
corresponding harmonics. As far as the scalar sum is also a scalar, the
result scalar field corresponding to (5) is defined as:
\begin{equation}
\phi(\vec{r},\tau)=\sum^{\infty}_{k=0}|q_k(x,y,\xi)|^2=
\sum^{\infty}_{k=0}g_k^2(\vec{r},\tau).
\label{n6}
\end{equation}
The important distinguish of the normalised field (5) from the scalar one
(6) is an oscillation, resonant nature of the first one. It is necessary
to note that the established correspondence between (5) and (6) has not
reciprocally a single meaning, because, for example, the information about
the mutual phases of harmonics is loosing. However, in considering only one
harmonic of the stable physical object function, this limitation is not
important. But the information about frequency will be still loosing. \\

\section{Quantum mechanics basic equations}
\label{sec:level2}

Omitting the corresponding indexes of $\psi$, $q$, $\omega$ for the notation
simplicity, we will represent the $k$-th harmonics of the spectrum expansion 
(5) as follows:
\begin{equation}
\psi(\vec{r},\tau)=\psi^b(\vec{r},\tau)\exp(i\omega\tau),~
\psi^b(\vec{r},\tau)=q(x,y,\xi)\exp(i\omega\eta).
\label{n7}
\end{equation}
Partial derivatives by $\vec{r}$ and $\tau$ for function $\psi^b$ will be
expressed in the following way:
\begin{equation}
{\partial\psi^b\over\partial\tau}=-\gamma\beta q_\xi e^{i\omega\eta}+
i\omega(\gamma-1)q e^{i\omega\eta},~
{\partial\psi^b\over\partial z}=\gamma q_\xi e^{i\omega\eta}-
i\gamma\omega\beta q e^{i\omega\eta},~
{\partial\psi^b\over\partial x(y)}= q_{x(y)} e^{i\omega\eta}.
\label{n8}
\end{equation}
It is designating here $f_s\equiv\partial f/\partial s$. 
For second derivatives the following expressions may be got: 
\begin{equation}
{\partial^2\psi^b\over\partial\tau^2}=\gamma^2\beta^2 q_{\xi\xi}
e^{i\omega\eta}-2i\gamma(\gamma-1)\omega\beta q_\xi e^{i\omega\eta}-
\omega^2(\gamma-1)^2 q e^{i\omega\eta},~
{\partial^2\psi^b\over\partial z^2}=\gamma^2 q e^{i\omega\eta}
(q_{\xi\xi}/q-\omega^2\beta^2)-
2i\gamma^2\omega\beta q_\xi e^{i\omega\eta}.
\label{n9}
\end{equation}
It is possible to cancel imaginary parts by combining the expressions for
the first derivative in time (8) and the second one in space (9). This
way for function $\psi^b(\vec{r},\tau)$ we will get the equation:
\begin{equation}
-i\gamma{1\over\psi^b}{\partial\psi^b\over\partial\tau}+
{1\over 2\omega\psi^b}\nabla^2\psi^b={1\over 2\omega}{\nabla^2q\over q}+
{\omega\over 2}(\gamma-1)^2.
\label{n10}
\end{equation}
Here $\nabla^2\equiv\partial^2/\partial x^2+\partial^2/\partial y^2+
\partial^2/\partial z^2$ is a Laplace operator.
The second term in the right part of the equation speeds to zero fast enough 
$(\sim\beta^4)$ with non-relativistic velocities. If it is possible to
separate the variables in some frame of references or at least to separate 
the time variable in the
function $q$, so $\nabla^2 q/q=u(x,y,z)$ and, supposing $\omega=mc/\hbar$,
where $m$ and $\hbar$ are some constants (usually $m$ is understood as
a rest mass, $\hbar$ is a Planck constant), we will get:
\begin{equation}
-i\hbar c\gamma{\partial\psi^b\over\partial\tau}+
{\hbar^2\over 2m}\nabla^2\psi^b=\left[{\hbar^2\over 2m}u(x,y,z)+
mc^2{(\gamma-1)^2\over 2}\right]\psi^b.
\label{n11}
\end{equation}
Designating $\hbar^2u(x,y,z)/2m=U(x,y,z)$, we will get in non-relativistic
case in passage to the limit $(\gamma\rightarrow 1)$ 
{\bf the Schr\H{o}dinger
equation} (conjugated to usually used) [7]. It is known that $U(x,y,z)$
has a meaning of the potential energy of the particle in the force field,
and $\psi^b(\vec{r},\tau)$ is agree with the de Broglie's description of
the particle wave properties.\\
The equation may be interpreted in the following way. The lengthy in space
physical object is changing by an external field, moving or changing of
the object internal structure will depend on this field. In distinguish
of the Schr\H{o}dinger equation, the equation (10) and, with 
some stipulations, (11) have to be true also in the relativistic case. 
It is necessary to pay attention to the particularity of the described 
passage to the limit to the potential function of the external forced field, 
because an interaction can be as complete as partial and also with new 
physical objects creation.\\
Combining the expressions analogous to (9) for function $\psi(\vec{r},\tau)$,
 it is possible to cancel imaginary terms in right part of the expressions. 
The following equation can be got:
\begin{equation}
{1\over\psi}\left({\partial^2\psi\over\partial\tau^2}-\nabla^2\psi\right)=
-\left({\nabla^2 q-\beta^2 q_{zz}\over q}+\omega^2\right).
\label{n12}
\end{equation}
There are included the functions only from space coordinates on the right 
part of the equation in the fundamental frame of references
$ (\nabla^2 q-\beta^2 q_{zz})/q=\nabla'^2 q'/q' $,
but on the left part - in frame $\bf K$. Thus, with the stability condition of
the considered physical object, it needs to be a scalar on the right part
in brackets. Designating this scalar as $(mc/\hbar)^2$, we will get {\bf the
Klein-Gordon-Fock equation} [8] for a free relativistic (pseudo-) scalar
particle with rest mass $m$, that corresponds to the conventional model,
when the plane monochromatic wave is confronted to the particle (without
spin). Thus, the introduced function (5) may be interpreted as a wave
function of the physical object.\\
The particular case of the zero scalar corresponds to {\bf the wave equation}.
In this particular case, real and imaginary parts of the functional
normalised field are the components of the electric and magnetic field
intensities, and the scalar field, corresponding to this functional is
a density distribution of the electromagnetic field. If the wave equation
is considered as a special case of the common description of the physical
objects, the introduced conceptions of the functional and corresponding
scalar fields are getting very interesting physics analogies.

\section*{conclusion}

Thus, the proposed model of the physical objects allowing to unify the
description of the corpuscular and wave properties of the real objects -
particles and fields, has a conceptual nature. The model correlates with
the existing quantum mechanics approach and, as follows, has an experimental
confirmation in non-relativistic and in a number of particular cases. Such
basic quantum mechanics equations as the Schr\H{o}dinger and 
Klein-Gordon-Fock
ones can be considered as a corollary from the proposed model properties
and properties of a space, in which the physical object exists (it is 
considered the Minkowski space in this paper). \\
The proposed model includes the absolute remote fields of the Minkowski
space in the description. On base of this possibility it is possible to
go further- inside the physical object and to try to make clear its
internal structure. So, this paper is supposed by the author as the first
one in a number of papers, continuing this theme, in direction of further
working up the theses following through the properties of the proposed model
of the physical object and the space properties. For example, there are
preparing papers about an internal structure of physical objects,
their interconnections. \\
This work is an effort to remove the conceptual internal contradictions
of the quantum mechanics theory. It is supposede, it will allow to find 
a way of creating in its frames the complete, correct, non-contradictory 
theory, dreamed by L. de Broglie, mentioned by M. Gell-Mann and "paved a 
way" by I. Prigogine.\\

% now the references. delete or change fake bibitem. delete next three
%   lines and directly read in your .bbl file if you use bibtex. 

% Put the label that you will use with \ref{} command in the braces 
% of the \label{} command. 
% \begin{figure}
% \caption{}
% \label{}
% \end{figure}

% that you will use with \ref{} command in the braces of the \label{} command 
% Insert the columnspecifiers (l, r, c, d, etc.) in the empty braces of the
% \begin{table}
% \caption{}
% \label{}
% \end{tabular}
% \end{table}

\end{document}